# Retail Market analysis in targeting sales based on Consumer Behaviour using Fuzzy Clustering - A Rule Based Model

D. Bhanu[#], Dr. S. Pavai Madeshwari[*]

[#]Department of Computer Applications, Sri Krishna College of Engineering and Technology, Coimbatore

[*]Department of Computer Science and Engineering, RMK Engineering College, Chennai

*Abstract—* **Product Bundling and offering products to customers is of critical importance in retail marketing. In general, product bundling and offering products to customers involves two main issues, namely identification of product taste according to demography and product evaluation and selection to increase sales. The former helps to identify, analyze and understand customer needs according to the demographical characteristics and correspondingly transform them into a set of specifications and offerings for people. The latter, concerns with how to determine the best product strategy and offerings for the customer in helping the retail market to improve their sales. Existing research has focused only on identifying patterns for a particular dataset and for a particular setting. This work aims to develop an explicit decision support for the retailers to improve their product segmentation for different settings based on the people characteristics and thereby promoting sales by efficient knowledge discovery from the existing sales and product records. The work presents a framework, which models an association relation mapping between the customers and the clusters of products they purchase in an existing location and helps in finding rules for a new location. The methodology is based on the integration of popular data mining approaches such as clustering and association rule mining. It focusses on the discovery of rules that vary according to the economic and demographic characteristics and concentrates on marketing of products based on the population.**

## I. INTRODUCTION

Product Bundling and offering products to customers is of critical importance in retail marketing. In this work, a predictive mining approach is presented that predicts sales for a new location based on the existing data. The major issue lies in the analysis of sales forecast based on the dependencies among the products and the different segment of customers, which helps to improve the market of the retail stores. The work presents a framework, which models an association relation mapping between the customers and the clusters of products they purchase in an existing location and helps in finding rules for an entirely new location. A novel methodology and model are proposed for accomplishing the task efficiently. The methodology is based on the integration of the popular data mining approaches such as clustering and association rule mining. It focusses on the discovery of rules that vary according to the economic and demographic characteristics and concentrates on marketing the products based on the population.

Today's manufacturers compete on price, adaptability and the variety of products which are all driven by customer satisfaction [1]. From a customer point of view, most customers purchase a product according to their individual preferences. Association rule mining is a significant data mining technique in generating rules based on the correlation between the sets of items that the customers purchase. To some extent, the rule generated gives us the dependencies among the set of customers and the products that they purchase. However, if the retail store plans to open a chain of stores at different location, the retailer has no idea about the customer attraction towards products at different locations. From the retailer's viewpoint, parameters / variables used to understand customer requirements are always fuzzy, uncertain and abstract. Researchers have analyzed the different aspects of the technique, but the shift in focus of increasing sales and customer targeting at different locations with the data available from a single point of sale is not much dealt with. It is without doubt that predicting sales at different locations with the data currently at hand need a solid methodology, which involves a large amount of data and knowledge inferred from the data. Therefore, the key element is to use the existing knowledge and patterns to generate new patterns or rules for different locations.

Association rules generated for a location at a point of sale cannot be effective in another location since the complete and complex behaviour of customers and their approach in selecting products are different. The relationship between the customer requirements and the practical supplies that the stores can provide to the customers are often not readily available for a different store at an entirely different location. Fortunately, the existing relationship or patterns turn out to be more specific, which depicts the relationship between the customers and the products they go for.

The major challenge of our work is the identification of rules for a different demographic population based on the existing records. Past research has not much focused upon the knowledge discovery of customer needs for a new point with the current available data. This paper proposes a novel approach in integrating the rule mining technique with the clustering approach to discover rules driven by customer needs for a typical new situation. As a result, the approach has the opportunity to take advantage of the wealth of the customer needs information and the alternative rules for the new situation, which better aids the retailer in designing his store layout.

## II. RELATED WORK

Since the introduction of the problem of mining association rules [1], several generate-and-test type of algorithms have been proposed for the task of discovering frequent sets. An efficient breadth-first or level-wise method for generating candidate sets, i.e., potentially frequent sets has been proposed in [3] [4] [5]. The method also called Apriori is the core of all known algorithms except the original one [1] and its variation for SQL, which have been shown inferior to the level-wise method. [4][5][7]

An alternative strategy for a database pass, using inverted structures and a general purpose DBMS, has been considered. Other work related to association rules includes the problem of mining rules with generalization [5], management of large amount of discovered rules and a theoretical analysis of an algorithm for a class of KDD problems including the discovery of frequent sets [7].A connectionist approach to mining rules is presented. All these approaches focus on generating rules, but a little attention is given to address the credibility of the rules outside of the dataset from which it was generated.

Association rule mining in general is used to express knowledge. Fuzzy set theory has been applied in many fields of data mining. Fuzzy clustering method is used mainly for improving the precision of results. Many fuzzy clustering methods have been proposed and used in the research literature. Fuzzy clustering with squared Minkowski distances have been proposed in [8]. Fuzzy c-means method was proposed by [6] for dynamic data mining. In [9], the authors cluster numerical dataset using relative proximity. In [11], the authors propose a methodology to cluster transactional data based on the distance metrics proposed by the K-means algorithm.

Most of the clustering methods are based on the Boolean logic model. They assume that the user's requirements can be characterized by the terms, which cannot be justified due to the fuzziness in data. The objective of the fuzzy clustering method is to partition the dataset into a set of clusters based on similarity [14]. In classical cluster analysis, each datum is assigned to exactly one cluster. Fuzzy clustering relaxes this constraint by allowing membership degrees, thus offering an opportunity to deal with data that belong to more than one cluster at the same time. The size, shape and various other parameters determine the extension of the cluster in different directions of the domain.

## III. PROBLEM FORMULATION

For a potential customer arriving the store, which customer group one should belong to according to customer needs, what are the preferred functional features or products that the customer focusses on and what kind of offers will satisfy the customer etc., finds to be the key factor in targeting customers to improve sales. Generally, a transactional database is created to record all the products purchased by the customer. To focus on the market segment that each customer falls into, the transaction database can be grouped into different clusters based on the customer needs.

In general, customer needs can be described as a set of features or attributes, $A = \{a_1, a_2...a_M\}$. Each attribute, $a_i \mid \forall i \in [1, 2... M]$, may take one out of a finite set of options, $A_i = \{a_{i1}, a_{i2},..., a_{in_i}\}$. Suppose all customers comprise a set, $C = (c_1, c_2,...,c_S)$, where 'S' denotes the total number of customers. The requirement information of a particular customer, $C_s \in C \mid \exists s \in [1... S]$, for the existing dataset can be depicted by a vector of certain options of the features. Hence, all customers can be grouped into a set, $A = \{\vec{a}_1, \vec{a}_2... \vec{a}_s\}$ which characterizes the customer domain. The functionality of each product is characterized by the set of functional requirements, $V = \{v_1, v_2... v_N\}$. Each functional requirement, $v_q = v_{qr} \mid \exists v_{qr} \in V_q$, where $r = 1,...,n_q$, denotes the rth possible value of $v_q$. Suppose, let us assume that all existing products comprise a set $P = \{p_1, p_2... p_T\}$, where T refers to the total number of products. The requirement specification for a particular product, $p_t \in P \mid \exists t \in [1, 2... T]$ can be represented as a vector of certain requirement values.

Similarly, all the instances of requirements in the product domain constitute a set, $V = \{\vec{v}_1, \vec{v}_2... \vec{v}_T\}$. The problem domain and the specification of the problem between the customer needs and the functional requirements are shown in Figure 1.

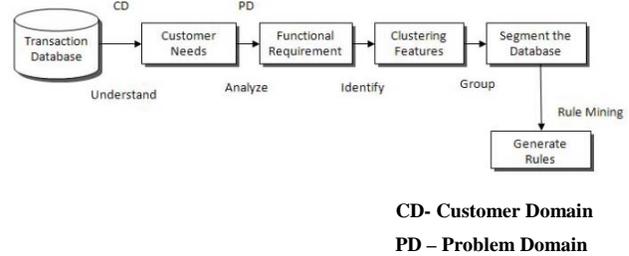

**CD- Customer Domain**

**PD – Problem Domain**

**Fig. 1 – Specification of problem domain**

The needs that exist in the customer requirements can be treated as the relationships between the feature vectors. Also, a study on customer behaviour indicates that the customers falling in the same category usually hold the same purchasing behaviour. While providing customer's interest based retail store layouts, the retailer must plan the customer needs. It is also seen that the customer falling into the same cluster say for eg., income, gender and age usually exhibits the same purchase behaviour and thus when opening a new store layout, the needs of the customer can be met by analyzing such clusters and offering them the products of their interest. Therefore, by analyzing the mean values of the different purchases of the different customers belonging to a cluster, the preference of an individual towards the purchase of a product can be easily found. Hence, the proposed methodology aims to generate rules for a new situation based on the rules that are obtained from the current knowledge base by applying fuzzy clustering means. Figure 2 shows the rule identification for a new setting based on Association Rule Mining and Fuzzy clusters.

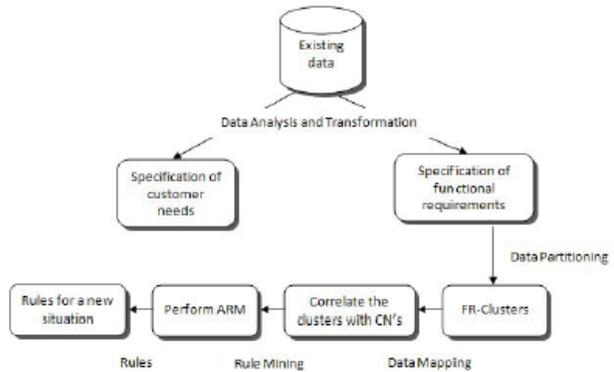

**Fig.2 – Rule identification for a new setting based on Association Rule Mining**

## IV. ARCHITECTURE OF THE PROPOSED SYSTEM

In order to obtain the association rules for a new store based on the analysis of customer transactions from the existing knowledge base, a Clustering based Association Rule Mining System architecture [CARMS] to predict sales at a different location is proposed. The system involves different consecutive stages communicating with one another in generating rules as data pre-processing and data partitioning, data transformation, association rule mining and presentation modules. Before proceeding to the rule mining of datasets, raw data must be pre-processed in order to be useful for knowledge discovery. Due to the uncertainty of customer requirements and their behaviour, it is necessary to pre-process the knowledge base. Based on the raw data stored in the knowledge base, target datasets should be identified, involving data cleaning and filtering tasks as integration of multiple databases, removal of noise, handling of missing data files etc.,

All target data should be organized into a usable transaction database. This involves the clear understanding of the variables, selection of attributes, which are more pertinent in generating rules for the new store. In the architecture proposed, the sales records and the product details are transformed into transaction data, which consists of a unique transaction identifier (TID). Transaction data consists of customer details and their affinity towards the products. Each customer is given a series of options on the selection of products based on the customer's attributes such as income, age and gender, which are recoded as the key operational features. The options of products that the customer desires can be stated as related Functional Requirements, which can serve as mandatory information for the predicting of sales at a new location. The result of the Customer Needs and the Functional Requirement mappings are also recorded in the transaction database.

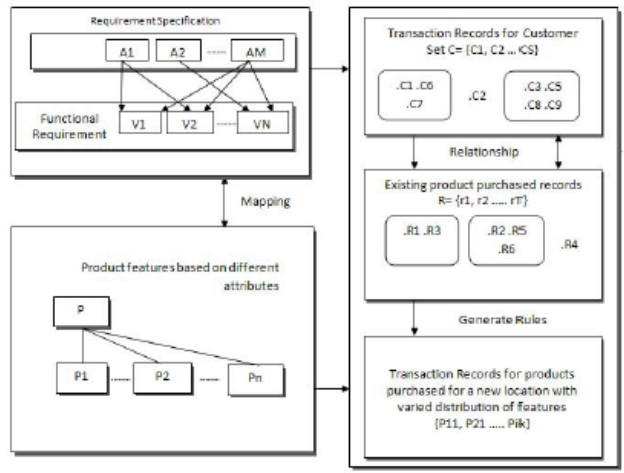

**Fig. 3 – The Block Diagram of CARMS**

The figure shown above in figure3 shows the Block Diagram of CARMS, which maps the customer needs and the Functional requirements and helps in finding rules for a new setting based on the existing transactional data.

**The Fuzzy Clustering Method for the formation of groups**

The behaviour of customers is often uncertain and vague and hence is frequently presented in linguistic forms. A concrete definition of customer behaviour is not possible and is a rare case. For the proposed architecture, the clustering method includes the following steps as distance measure and fuzzy clustering. As the preparatory stage for fuzzy clustering, the distance measure module measures the dissimilarity between the functional requirement instances in order to find the fuzzy compatible relations among such data objects.

**Data Pre-processing**

Many data types are available in fuzzy clustering, such as binary variables, ordinal variables, ratio-scaled variables and even a mix of these variables. Because of the influences in the demography at different locations, data standardization should also be considered. The data standardization for numerical features alone is specified for the current discussion, as all the other data types can be converted to the numerical data type. Let $F_i = (f_{i1}, f_{i2}...f_{iM}) | \forall i \in [1, 2..., T]$ be the transaction record, where M denotes the product features. A standard deviation equation is applied on the raw data to standardize the data.

Suppose $f'_{ik} = f_{ik} - f'_k / S_k$ ($\forall i \in [1,2,...,T], \forall k = 1,2,...,M$), where $f_{ik}$ corresponds to the option value of the ith transaction record for the specification $f_k$, where $f'_k = 1/T \sum_{i=1}^{T} f_{ik}$. It denotes the average value of all the transactions as to the kth specification, where $S_k = \sqrt{1/T \sum_{i=1}^{T} | f_{ik} - f_k^2 |^2}$. Unlike the max-min standardization, it cannot transform the raw data to values between 0 and 1, but eliminate the influence of different dimensions. Figure4 shows the method for pre-processing for customer need identification and Figure 5 shows the pre-processing for product requirement need identification.

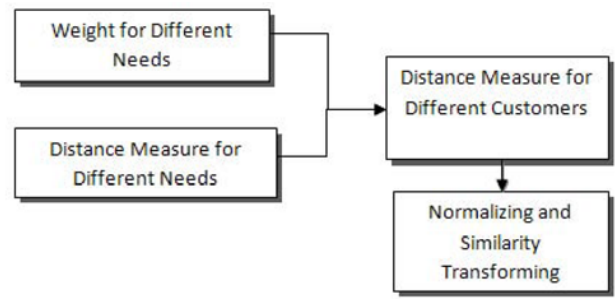

**Fig. 4 - Pre-processing for Customer Need Identification**

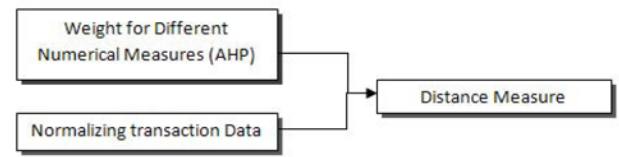

**Fig.5 - Pre-processing for Requirement Need Identification**

**Prioritization of variables by assigning weights**

The specification of Functional requirements involves multiple variables, i.e., $V = \{v_q | \forall q = 1... N\}$. These variables constitute to the overall functionality of a product purchase, which is unique in its own form. Hence, Functional requirements variables should be prioritized to differentiate their effects. The relative importance of requirements variables is usually quantified by assigning different weights. That is, each $v_q$ is associated with a weight, $w_q$, subjective to $\sum_{q=1}^{N} w_q = 1$. For the proposed architecture, AHP [20] is adopted for prioritization of functional requirements variables, owing to its advantages in maintaining consistence among a large number of variables through pair-wise comparisons.

**Computation of Distance to find the similarity measures**

The Functional Requirement clustering includes two steps as distance measure and fuzzy clustering. As the first stage, the distance measure module measures the dissimilarity between FR instances in order to define the fuzzy compatible relations among data objects.

Distance measure: In general, each FR instance $v'_t = [v_{1t}, v_{2t} \ldots v_{qt}, \ldots, v_{Nt}] \in v'$, where $\forall v_{qt} = v_{qr}, \exists v_{qr} \in V_q, \forall r = 1,\ldots,n_q$, may involve three types of variables: numerical, binary and nominal variables. For example, $v_{1t}$ may be a numerical value while $v_{2t}$ may be a binary or nominal value. The distance between any two instances means the dissimilarity between them and thus is measured as a composite distance of the distance components corresponding to these three types of variables.

Numerical Measures – a number of methods of distance measure have been proposed for purpose of numerical clustering, including the Euclidean distance, Manhattan distance, Minkowski distance and weighted Euclidean distance measure [13]. The proposed architecture employs the weighted Euclidean distance. It is computed as following:

$d_{numerical}(v'_i, v'_j) = \sqrt{\sum_{q=1}^{Q}(w_q(N\_v'_{qi} - N\_v'_{qj}))^2}$, where

$d_{numerical}(v'_i, v'_j)$ indicates the numerical distance between two FR instances, $v'_i$ and $v'_j$, where $\forall v'_i, v'_j \in V$, $w_q$ means the relative importance of the q-th numerical requirement variable, $V_q \in V^{numerical} \subseteq V$, Q represents the total number of numerical FR variables among the total size – N FR variables ($Q \leq N$) and $N\_v_{qt}$ denote the normalized values $V_{qi}$ and $V_{qj}$ respectively.

Binary FRs – A binary variable assumes only two states: 0 or 1, where 0 means the variable is absent and 1 means it is present. The architecture uses a well –accepted coefficient for assessing the distance between symmetric binary variables, called the simple matching coefficient [16]. It is calculated as the following,

$d_{binary}(v'_i, v'_j) = \alpha_2 + \alpha_3 / \alpha_1 + \alpha_2 + \alpha_3 + \alpha_4$, where

$d_{binary}(v'_i, v'_j)$ indicates the binary distance between two FR instances, $v'_i$ and $v'_j$, $\forall v'_i, v'_j \in V$, $\alpha_1$ is the total binary variables in V that equal to 1 for both $v'_i$ and $v'_j$, $\forall v'_i, v'_j \in V$, $\alpha_2$ is the total binary variables in V that equal to 1 for $v'_i$ but 0 for $v'_j$, $\alpha_3$ is the total binary variables in V that equal to 0 for $v'_i$ but 1 for $v'_j$, and $\alpha_4$ is the total number of binary FR variables that equal to 0 for both $v'_i$ and $v'_j$.

Nominal FRs – A nominal variable can be regarded as a generalization of binary variable in that it takes on more than two states. This type of variables cannot be expressed by numerical values but by qualitative expression with more than one option.

Therefore the simple matching coefficient can be used here to measure the nominal distance between two FR instances containing nominal FR variables [16].

$d_{nominal}(v'_i, v'_j) = \beta - \gamma / \beta$, where $d_{nominal}(v'_i, v'_j)$ indicates the nominal distance between two FR instances, $v'_i$ and $v'_j$, $\forall v'_i, v'_j \in V$, $\gamma$ means the total number of nominal FR variables in V (i.e., $v_q \in V^{nominal} \subseteq V$) that assume the same states for $v'_i$ and $v'_j$ and $\beta$ is the total number of nominal variables among total size – N FR variables ($\beta \leq N$). Given a set of functional requirement variables, $V = \{v_1, v_2 \ldots v_N\}$, every functional requirement instance assumes a certain value for each variable, and thus consists of a combination of numerical, binary and /or nominal requirement values, that $V^{numerical} \cup V^{binary} \cup V^{nominal} = V$. As a result, the overall distance between $v'_i$ and $v'_j$ comprises three components: the numerical, binary and the nominal distances. A composite distance can thus be obtained by the weighted sum: $d(v'_i, v'_j) = W_{numerical} d_{numerical}(v'_i, v'_j) + W_{binary} d_{binary}(v'_i, v'_j) + W_{nominal} d_{nominal}(v'_i, v'_j)$, $\sum (W_{numerical} + W_{binary} + W_{nominal}) = 1$, where $W_{numerical}$, $W_{binary}$ and $W_{nominal}$ refer to the relative importance of numerical, binary and nominal distances respectively. These weights can be determined in the similar way as that of requirement variables applying the AHP.

**Fuzzy clustering**

Three main methods are available to facilitate fuzzy clustering, including fuzzy netting graph, maximum generated graph and transitive closure method. The former two methods can be implemented through the fuzzy compatible matrix directly, while the latter methods need to convert fuzzy compatible matrix to fuzzy equivalent matrix. It is then transformed into a Boolean equivalent matrix. It means that besides the fuzzy compatible relation that is composed of both symmetric and reflexive relations, the fuzzy equivalent relation utilized to construct the fuzzy equivalent matrix should have the transitive relation.

Here, we adopt the Boolean equivalent matrix to obtain the clusters of customer needs and product features, respectively. For the customer transaction records $C_i$, R is transitive iff $R \circ R \leq R$ ($\Leftrightarrow \vee_{k=1}^{T}(\rho(C_i, C_k) \wedge \rho(C_k, C_i))$. Similarly, for the transaction record $F_i$, G is transitive iff $G \circ G \leq G$. The process is performed by the max-min transitive closure for R and G respectively. This method is based on the following model: $t(R) = \bigcup_{k=1}^{t} R^k$, where t (R) is transitive closure and $R^k$ denotes the max-min operation for k items. The above model implies that the transitive closure $t(R)$ can be obtained by max-min operation for most T times. In this way, the transitive relation is satisfied and the fuzzy compatible matrix $R_{T \times T}$ and $G_{T \times T}$ are converted into the fuzzy equivalent matrix t(R) and t (G) respectively.

The following step is to transform the fuzzy equivalent matrix to the Boolean equivalent one. Assuming $t(R) = [\rho'(C_i, C_j) | \forall (C_i, C_j) \in T \times T]$, the $\alpha$-cut of $t(R)$ is represented as $R_\alpha = [v(C_i, C_j)] T \times T$, where $v(C_i, C_j) = 1$, if $\rho'(C_i, C_j) \geq \alpha$; otherwise $\upsilon(c_i, c_j) = 0$. It is obvious that Boolean matrix $R_\alpha$ is an equivalent matrix too. If all the elements of two rows in $R_\alpha$ are identical, the transaction records corresponding to the rows are classified into the same cluster. The clusters are described as $U_T / R_\alpha = \{g_1, g_2 \ldots g_s\}$, where $U_T$ is the universe of all the transaction records and s is the number of clusters. Similarly, we can derive the clusters with respect to product specifications by setting different threshold values for $\beta$. The clusters are depicted as $U_T / G_\beta = \{p_1, p_2, \ldots, p_l\}$. Because the equivalent matrix $R_\alpha$ and $G_\beta$ are defined by a corresponding binary equivalent relation, they can be also equally viewed as the equivalent relation. Figure 6 shows the Fuzzy clustering method for grouping customers by converting the customer needs and functional requirements onto a fuzzy equivalent matrix.

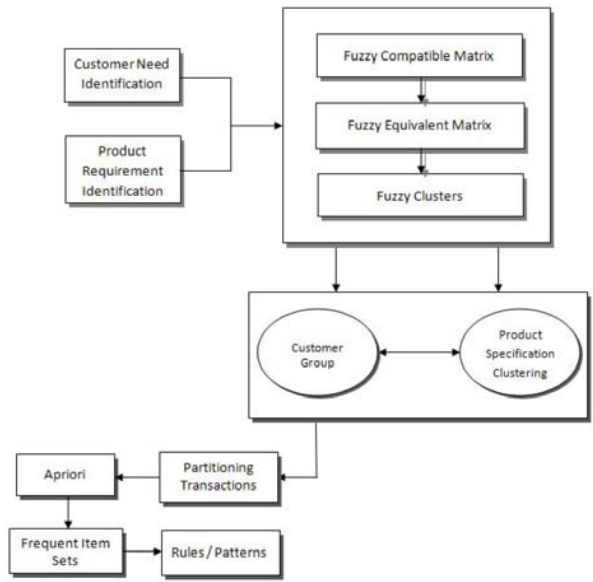

**Fig. 6 – Fuzzy Clustering**

**Rule Evaluation**

The Apriori algorithm is the most well known association rule-mining algorithm. At first, we give the following definitions [10]:
Definition 1 : Given a set of items I = { $I_1, I_2, \ldots, I_s$ }, and the database of transaction records D = { $t_1, t_2, \ldots, t_n$ }, where $t_1 = \{ I_{i1}, I_{i2}, \ldots, I_{ik} \}$ and $I_{ij} \in I$, an association rule is an implication of the form $X \Rightarrow Y$ where X, Y $\subset$ I and X $\cap$ Y = $\Phi$.
Definition 2: The support (s) for an association rule $X \Rightarrow Y$ is the percentage of transactions in the database that contain $X \cup Y$. That is, support ($X \Rightarrow Y$) = P ($X \cup Y$), P is the probability.

Definition 3: The confidence or strength ($\Phi$) for an association rule ($X \Rightarrow Y$) is the ratio of the number of transactions that contain $X \cup Y$ to the number of transactions that contains X. That is confidence ($X \Rightarrow Y$) = P (Y|X). The algorithm uses the following property: If an itemset satisfies the minimum support threshold, so do all its subsets. The key of Apriori algorithm is to generate the large itemsets and then to generate association rules. The details of the algorithm are more specifically given in [10].

The unions of transaction records in the clusters that make the dependency maximum are often more representative than other transaction ones. Therefore, we can partition the target transaction table with them to decrease the scale of data mining without loss of the information content. In general, the focus must be more on the cluster groups than the individual customers, since the groups can reflect the characteristics of individual customers. Based on the previous considerations, the rule mining process can be divided into three steps:

Choosing the suitable product clusters, which make the dependency maximum; calculating the mean values and the corresponding variation as to specification for each transaction in the cluster.

Choosing the unions of transaction records of product specifications in the clusters that make the dependency maximum replacing items in the chosen transactions with the new items represented with mean value and variation range. Supposing X represents the items of product features and Y represents the items of the requirement alternatives in the same transaction record, by implementing the Apriori algorithm with minimum support (s) and confidence ($\phi$), the association rule $X \Rightarrow Y$ depicts the relation between the product specifications and the requirements alternative. Figure 7 show the overall architecture of CARMS.

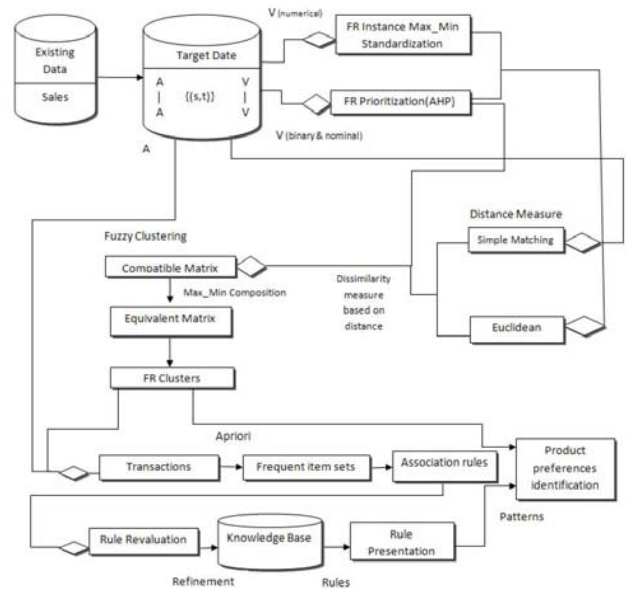

**Fig. 7 – CARMS overall architecture**

## V. CASE STUDY

The proposed architecture has been tested for a retail store, which has its branches at more than 10 locations. The store has a large variety of different products to be offered for the customer. The practical problem is that the retailer is confused with the store layout and the items to be put on sale at a different location when he opens a new store at an entirely new location. Since customers at different age groups, gender and with different income levels have distinct needs in procuring things from the store, a new strategy has to be developed for the new store with the study of the existing demographic study at the new location. In the existing methodology, some approaches are used to discover the potential relationships between the customer groups and the products they purchase, the extension of predicting sales at an entirely new location is not predicted. Therefore, the methodologies seem to be impractical from the new store perspective. In order to predict sales for a new location, the customers with similar preferences are grouped into the same category and the properties associated with an individual customer are shifted towards the total group.

The dependency between the group behaviour and their purchase of items plays a key factor in generating association rules. According to the marketing survey, irrespective of the customer category, the main concern of people towards the product can be summarized as in Table 1. The features expected on the product requirements and their options are shown in Table 2. Among the features, some are interval-based variables, which are specified in ranges. Based on the existing records, target data are identified and organized into a transaction database as shown in Table 3. The target transaction database is obtained from the existing database. For illustrative simplicity, only 15 records were considered from the transaction database. As shown in Table 3, each customer order indicates the customer's choice of certain feature options. The AHP is implemented to prioritize the nine features and we get the relative vector

$\vec{W}$ = {0.115+0.351+.067+.034+.021+.075+.146+.083+.106}

Due to the different metrics used for the functional requirement variables, all the FR instances in Table 3 are standardized based on the max-min standardization. The results of the distance measures for the binary and numerical requirement instances are shown in Figures 10 and 11 respectively. Based on the max-min standardization and relative weights, the dissimilarity matrix is obtained as shown in Figure 8. Subsequently, the weighted Euclidean distance is used to get the dissimilarity matrix and the fuzzy clustering module is adopted to obtain the fuzzy equivalent matrix as shown in Figure.9. By setting different similarity thresholds for the fuzzy equivalent matrices, Boolean equivalent matrices can be obtained.

**Table 1: List of Customer Needs**

| Feature | Description | Options |
|---|---|---|
| $a_1$ | Safe keeping | $a_{11}$ – High |
| | | $a_{12}$ - Low |
| $a_2$ | Resilience | $a_{21}$ – High |
| | | $a_{22}$- Medium |
| | | $a_{23}$ – Low |
| $a_3$ | User Satisfaction | $a_{31}$ – High |
| | | $a_{32}$ – Medium |
| | | $a_{33}$ - Low |
| $a_4$ | Strength | $a_{41}$ – High |
| | | $a_{42}$ - Low |
| $a_5$ | Cost | $a_{51}$ – High |
| | | $a_{52}$ – Medium |
| | | $a_{53}$ - Low |
| $a_6$ | Ease of Use | $a_{61}$ – High |
| | | $a_{62}$ – Medium |
| | | $a_{63}$ - Low |

**Table 2: List of Functional Requirements**

| Feature | Description | Options |
|---|---|---|
| v1 | Frequency of Purchase | $v_{11}$ – Once in 2 weeks |
| | | $v_{12}$ – Once in 4 weeks |
| | | $v_{13}$ – Once in 8 weeks |
| v2 | Cost | $v_{21}$ – Expensive |
| | | $v_{22}$ – Medium |
| | | $v_{23}$ – Less expensive |
| v3 | Performance | $v_{31}$ – Maximum |
| | | $v_{32}$ – Medium |
| | | $v_{33}$ - Low |
| v4 | Size | $v_{41}$ – Large |
| | | $v_{42}$ – medium |
| | | $v_{43}$ – Low |
| v5 | Age | $v_{51}$ – 0-20 |
| | | $v_{52}$ – 21-40 |
| | | $v_{53}$ – 41-60 |
| v6 | Gender | $v_{61}$ – Male |
| | | $v_{62}$ - Female |
| v7 | Income | $v_{71}$ – < 15000 |
| | | $v_{72}$ – 16000 – 30000 |
| | | $v_{73}$ – 31000-50000 |
| v8 | Usage | $v_{81}$ – High |
| | | $v_{82}$ – Medium |
| | | $v_{83}$ - Low |
| v9 | Manifestation | $v_{91}$ – yes |
| | | $v_{92}$ - No |

**Table 3: Transaction Database**

| Transaction Records (TID) | Customer Needs | Functional Requirements |
|---|---|---|
| T001 | a11,a21,a31, a44,a51,a62 | v11,v21,v31,v42, v52,v62,v72,v81,v92 |
| T002 | a11,a21,a43, a51 | v11,v21,v31,v41,v51, v62,v72,v81 |
| T003 | a12,a22, | v12,v21,v33,v43,v52, |
|  | a33,a62 | v61,v72,v83,v91 |
| … | … | … |
| T014 | a13,a21,a32, a42,a63 | v12,v22,v43,v52, v62,v72,v81,v91 |
| T015 | a12,a21,a32, a41,a61 | v11,v22,v31,v42, v53,v71,v82 |

**Figure 8: Dissimilarity matrix based on distance measures**

**Fig.9: Fuzzy Equivalent Matrix**

**Fig. 10: Result of distance measures for binary requirement instances**

**Fig.11: Result of distance measures for numerical requirement Instances**

Figure 12 shows the result of mapping between the customer groups and product clusters in order to get the customer groups that have the maximum dependency to clusters of product features. The characteristics of each cluster involves the specification of a set of base values together with the related variation ranges, and therefore can be used to suggest standard settings for a new location. The items are then added to the transaction database. The link of each customer preferences is then linked to the corresponding cluster that the customer belongs to. All data that are recorded in the transaction database is fed as input for the Apriori algorithm which generates rules based on the support and confidence measures. The output is guaranteed such that only those rules with the highest values for the specified measures are found according to user-defined threshold settings. Figure 13 shows the screenshot of the implemented association rule-mining algorithm. At the end of mining, the system generates 35 rules, based on the specified support and confidence.

The chosen association rules

1. If Age = 35 and Gender = M $\Rightarrow$ beer and cigarettes is common ( Support = 0.4,Confidence = 0.6)
2. If Age = 15 and Gender = M $\Rightarrow$ Coke and Chips is common ( Support = 0.4,Confidence = 0.6)
3. If Age = 35 and Gender = F $\Rightarrow$ milk and sugar is common ( Support = 0.4,Confidence = 0.6)
4. If Age = 15 and Gender = F $\Rightarrow$ Ice-cream and chocolates is common ( Support = 0.4,Confidence = 0.6)

……

In terms of the results of the customer clustering, we can see that the transactions with similar preferred customer needs are clustered into the same class. If for a new potential customer, things are to be planned, then, we can place the new customer into the corresponding customer group based on the mean value of the clusters obtained.

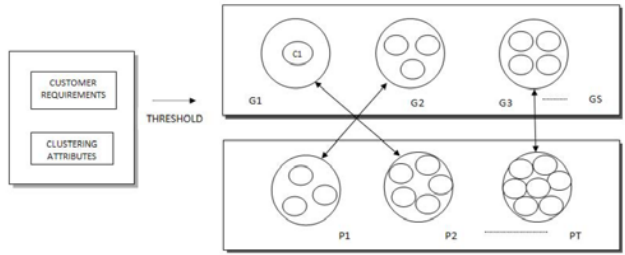

**Fig.12: Mapping between customer needs and clusters**

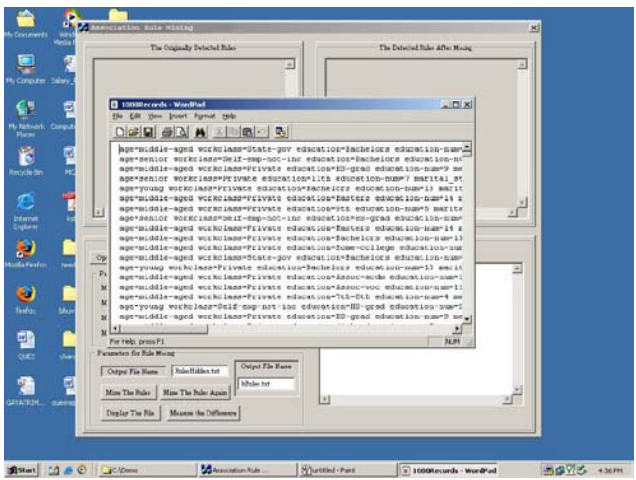

**Fig.13 – The results after Mining**

## VI. CONCLUSION

In this work, an efficient architecture is proposed to discover customer group-based rules if a retailer want to open his outlet at an entirely new location. In order to obtain the rules, both the customer and the product domains are bridged based on fuzzy clustering. Association rule mining and Fuzzy clustering were incorporated to analyze the similarity between customer groups and their preferences for products. The complete set of rules generated can be stored in a separate knowledge base. Then, for the stated or required customer needs, we can categorize the corresponding customer groups and can find the clusters to which the customer belongs. Finally, with the different options that the customer would prefer upon, we can predict the layouts and the items for the new store.